\documentclass[11pt,american,onecolumn,nofonttune]{IEEEtran}
\usepackage[T1]{fontenc}
\usepackage[latin9]{inputenc}
\usepackage[letterpaper]{geometry}
\usepackage{color}
\usepackage{amsmath}
\usepackage{amssymb}
\usepackage{esint}

\makeatletter
\usepackage{cite}

\newcommand{\e}{\mbox{e}}
\renewcommand{\d}{\mbox{d}}

\let\newcommand=\providecommand

\newcommand{\sba}[2]{\raisebox{-1mm}{\hbox to 0pt{\hspace*{-4mm}\resizebox*{#1mm}{#2mm}{\includegraphics{barra.epsi}}\hss}}}

\makeatother

\usepackage{babel}

\begin{document}

\title{An Alternative Derivation of Johannisson's Regular Perturbation Model}

\author{A. Bononi, P. Serena {July 19, 2012} \\Dip. Ingegneria Informazione,
Università degli Studi di Parma, Parma, Italy}
\maketitle
\begin{abstract}
We provide here an alternative derivation of the generalization of
the nonlinear Turin model for dispersion unmanaged coherent optical
links provided in Johannisson's report \cite{Pontus}. \end{abstract}

\section{Introduction}

Goal of this paper is to provide a simplified derivation of the results
appearing in the recent ArXiv posting of P. Johannisson \cite{Pontus}
on a generalization of the well-known frequency-domain nonlinear interference
(NLI) analytical model for dispersion unmanaged (DU) coherent systems
introduced by Turin's group in \cite{carena_JLT}.

\section{Frequency Domain NLI RP1 Solution}

We start from the dual-polarization (DP) single-channel first-order
Regular Perturbation (RP1) solution of the dispersion-managed nonlinear
Schroedinger equation (DMNLSE) (\cite{Bononi_OE}, Appendix 2):

\begin{equation}
\tilde{\boldsymbol{U}}(L,f)=\tilde{\boldsymbol{U}}(0,f)+\tilde{\boldsymbol{U}}_{p}(L,f)\label{eq:RP1-1}\end{equation}
where $L$ is the total link length, and the NLI \emph{p}erturbation
field is \begin{equation}
\tilde{\boldsymbol{U}}_{p}(L,f)=-jP_{0}\iint_{-\infty}^{\infty}\mathcal{K}(f_{1}f_{2})\tilde{\boldsymbol{U}}(0,f+f_{1})\tilde{\boldsymbol{U}}^{\dagger}(0,f+f_{1}+f_{2})\tilde{\boldsymbol{U}}(0,f+f_{2})\d f_{1}\d f_{2}\label{eq:RP1single}\end{equation}
where:

i) boldface fields are 2x1 vectors containing the Fourier transforms
(denoted by a tilde) of the X and Y polarizations in the transmitter
polarization frame of reference; a dagger stands for transposition
and conjugation; and the DP field power is normalized to an arbitrary
reference power $P_{0}$ (which in \cite{Pontus} is chosen as the
\emph{per-polarization} average power);

ii) the un-normalized scalar frequency kernel is defined as (Cfr.
\cite{Pontus} eq. (45) and \cite{Bononi_OE} eq. (25)): \begin{equation}
\mathcal{K}(F)\triangleq\int_{0}^{L}\gamma'(s)\mathcal{G}(s)\e^{-jC(s)(2\pi)^{2}F}ds\label{eq:DMkernel}\end{equation}
where $F=f_{1}f_{2}$ is the product of two frequencies, $\gamma'=\frac{8}{9}\gamma$
with $\gamma$ the fiber nonlinear coefficient, $\mathcal{G}(s)$
the line power gain from $z=0$ to $z=s$, and $C(s)=-\int_{0}^{s}\beta_{2}(z)dz$
is the cumulated dispersion in the transmission fibers (with dispersion
coefficient $\beta_{2}$) up to coordinate $s$. We choose here the
frequency-normalizing rate in \cite{Bononi_OE} as $R=1$, i.e., we
do not normalize the frequency axis, as in \cite{Pontus}. In \cite{Bononi_OE}
we use the normalized kernel\[
\tilde{\eta}(F)=\frac{\mathcal{K}(F)}{\mathcal{K}(0)}\]
which then at $F=0$ equals 1. The nonlinear phase referred to power
$P_{0}$ is \[
\Phi_{NL}=P_{0}\mathcal{K}(0).\]

Hence by multiplying and dividing by $\mathcal{K}(0)$ we can recast
(\ref{eq:RP1single}) as\begin{equation}
\tilde{\boldsymbol{U}}_{p}(L,f)=-j\Phi_{NL}\iint_{-\infty}^{\infty}\tilde{\eta}(f_{1}f_{2})\tilde{\boldsymbol{U}}(0,f+f_{1})\tilde{\boldsymbol{U}}^{\dagger}(0,f+f_{1}+f_{2})\tilde{\boldsymbol{U}}(0,f+f_{2})\d f_{1}\d f_{2}\label{eq:RP1single-1}\end{equation}
 From (\ref{eq:RP1single-1}), the X component of the RP1 solution
writes explicitly as \begin{align}
\frac{\tilde{U}_{x,p}(L,f)}{-j\Phi_{NL}}= & \iint_{-\infty}^{\infty}\tilde{\eta}(f_{1}f_{2})\tilde{U}_{x}(0,f+f_{1})\tilde{U}_{x}^{*}(0,f+f_{1}+f_{2})\tilde{U}_{x}(0,f+f_{2})\d f_{1}\d f_{2}+\nonumber \\
 & \,\iint_{-\infty}^{\infty}\tilde{\eta}(f_{1}f_{2})\tilde{U}_{x}(0,f+f_{1})\tilde{U}_{y}^{*}(0,f+f_{1}+f_{2})\tilde{U}_{y}(0,f+f_{2})\d f_{1}\d f_{2}\label{eq:field_x}\end{align}
where the first line gives the self-phase modulation (SPM) of X on
X, while the second line gives the intra-channel cross-polarization
modulation (I-XPolM) of Y on X. A perfectly dual expression for component
Y is obtained by exchanging the indices $x$ and $y$.

\section{Gaussian Assumption and Johannisson's Result}

In \cite{Pontus},\cite{carena_JLT} the key assumption is that the
input fields are composed of independent spectral lines with \emph{Gaussian}
amplitude:\begin{align*}
\tilde{U}_{x}(0,f) & =\sqrt{f_{0}}\sum_{k=-\infty}^{\infty}\xi_{k}\sqrt{\hat{G}_{x}(kf_{0})}\delta(f-kf_{0})\\
\tilde{U}_{y}(0,f) & =\sqrt{f_{0}}\sum_{k=-\infty}^{\infty}\zeta_{k}\sqrt{\hat{G}_{y}(kf_{0})}\delta(f-kf_{0})\end{align*}
with $\xi_{k}$ and $\zeta_{k}$ independent identically distributed
standard (i.e. zero-mean unit variance) circular complex Gaussian
random variables (RV). Such signals do have a \emph{per-polarization}
power spectral density $\hat{G}_{x/y}(f)$ (normalized to $P_{0}$)
in the limit $f_{0}\to0$ \cite{carena_JLT}. Then after long statistical
averaging calculations, one gets the power spectral density of the
$\tilde{U}_{x,p}(L,f)$ RV as (\cite{Pontus}, eq. (89). Note that
our PSD $\hat{G}(f)$ is normalized such that $G(f)\equiv P_{0}\hat{G}(f)$,
where $G$ is the un-normalized PSD per polarization. Also, $P_{x}=P_{0}\int_{-\infty}^{\infty}\hat{G}_{x}(f)df$
and $P_{y}=P_{0}\int_{-\infty}^{\infty}\hat{G}_{y}(f)df$)\begin{align}
\hat{G}_{x,p}(f)= & P_{0}^{2}\{2\iint_{-\infty}^{\infty}\left|\mathcal{K}((f_{1}-f)(f_{2}-f))\right|^{2}\hat{G}_{x}(f_{1})\hat{G}_{x}(f_{2})\hat{G}_{x}(f_{1}+f_{2}-f)\d f_{1}\d f_{2}\nonumber \\
+ & \iint_{-\infty}^{\infty}\left|\mathcal{K}((f_{1}-f)(f_{2}-f))\right|^{2}\hat{G}_{x}(f_{1})\hat{G}_{y}(f_{2})\hat{G}_{y}(f_{1}+f_{2}-f)\d f_{1}\d f_{2}\nonumber \\
+ & \mathcal{K}(0)^{2}\hat{G}_{x}(f)\left(4(\int_{-\infty}^{\infty}\hat{G}_{x}(f)df)^{2}+4\int_{-\infty}^{\infty}\hat{G}_{x}(f)df\int_{-\infty}^{\infty}\hat{G}_{y}(f)df+(\int_{-\infty}^{\infty}\hat{G}_{y}(f)df)^{2}\right)\}\label{eq:G_xp}\end{align}
and a dual expression for Y is obtained by swapping $x\leftrightarrow y$.
Recall that $\hat{G}_{x,p}(f)$ is the NLI PSD, normalized by $P_{0}$. 

An equivalent form of (\ref{eq:G_xp}) is the following\begin{align}
\hat{G}_{x,p}(f)= & \Phi_{NL}^{2}\{2\iint_{-\infty}^{\infty}\left|\tilde{\eta}(f_{1}f_{2})\right|^{2}\hat{G}_{x}(f+f_{1})\hat{G}_{x}(f+f_{2})\hat{G}_{x}(f+f_{1}+f_{2})\d f_{1}\d f_{2}\nonumber \\
+ & \iint_{-\infty}^{\infty}\left|\tilde{\eta}(f_{1}f_{2})\right|^{2}\hat{G}_{x}(f+f_{1})\hat{G}_{y}(f+f_{2})\hat{G}_{y}(f+f_{1}+f_{2})\d f_{1}\d f_{2}\nonumber \\
+ & \hat{G}_{x}(f)\left(4(\int_{-\infty}^{\infty}\hat{G}_{x}(f)df)^{2}+4\int_{-\infty}^{\infty}\hat{G}_{x}(f)df\int_{-\infty}^{\infty}\hat{G}_{y}(f)df+(\int_{-\infty}^{\infty}\hat{G}_{y}(f)df)^{2}\right)\}\label{eq:G_xp-2}\end{align}
which better shows the formal parallel with the field equation (\ref{eq:field_x}):
the field double integral in $f_{1},f_{2}$ of the product kernel-field-field$^{*}$-field
becomes a PSD double integral in $f_{1},f_{2}$ of the product squared
kernel magnitude-PSD-PSD-PSD. 

It is the purpose of the remaining part of this paper to provide a
new proof of (\ref{eq:G_xp-2}).

\section{The New Proof}

We now start from (\ref{eq:field_x}) and make the following two assumptions
regarding the input X,Y fields $U_{x}(0,t)$, $U_{y}(0,t)$:

1) they are wide-sense stationary (WSS);

2) they are jointly Gaussian processes.\\

Regarding assumption 1), we plan to exploit the following extension
of result (\cite{papo}, p. 418, eq. (12-76)~):\\

\textbf{\emph{Theorem}}\textbf{ 1}

Consider the jointly WSS stochastic processes $x(t)$ and $y(t)$,
and let \[
\tilde{X}(f)\equiv\mathcal{F}[x(t)]=\int_{-\infty}^{\infty}x(t)\e^{-j2\pi ft}\d t\]
 the Fourier transform of $x$ ( in the mean-square (MS) sense), and
$\tilde{Y}(f)$ is similarly defined. Let their cross power spectral
density (PSD) be $G_{xy}(f)=\mathcal{F}[R_{xy}(\tau)]=\mathcal{F}[E[x(t+\tau)y^{*}(t)]]$.
Then\begin{equation}
E[\tilde{X}(f)\tilde{Y}^{*}(u)]=G_{xy}(f)\delta(f-u)\equiv G_{xy}(f)\delta(u-f)\qquad\square\label{eq:papoulis}\end{equation}
\\

As a byproduct, we also have \[
E[\tilde{X}(f)\tilde{X}^{*}(u)]=G_{x}(f)\delta(f-u)\equiv G_{x}(f)\delta(u-f).\]

This theorem thus shows that the Fourier transform of \emph{any MS-integrable
WSS process} is nonstationary white noise, and thus \emph{the spectral
lines of its Fourier transform are uncorrelated}.

Regarding assumption 2), we plan to exploit the following result,
known as the \emph{complex Gaussian moment theorem} (CGMT), a generalization
to complex variables of Isserlis theorem \cite{reed,goodman}:\\

\textbf{\emph{Theorem 2}}

Let $U_{1},U_{2},\ldots,U_{2k}$ be zero-mean jointly circular complex
Gaussian random variables. Then\begin{equation}
E[U_{1}^{*}U_{2}^{*}...U_{k}^{*}U_{k+1}U_{k+2}...U_{2k}]=\sum_{\pi}E[U_{1}^{*}U_{p}]E[U_{2}^{*}U_{q}].....E[U_{k}^{*}U_{r}]\label{eq:isserlis}\end{equation}
where $\sum_{\pi}$ denotes a summation over the $k!$ possible permutations
$(p,q,...,r)$ of indices $(k+1,k+2,...,2k)\qquad\square$\\

For instance, \begin{align}
E[U_{1}^{*}U_{2}^{*}U_{3}^{*}U_{4}U_{5}U_{6}] & =E[U_{1}^{*}U_{4}]E[U_{2}^{*}U_{5}]E[U_{3}^{*}U_{6}]\nonumber \\
 & +E[U_{1}^{*}U_{4}]E[U_{2}^{*}U_{6}]E[U_{3}^{*}U_{5}]\nonumber \\
 & +E[U_{1}^{*}U_{5}]E[U_{2}^{*}U_{4}]E[U_{3}^{*}U_{6}]\nonumber \\
 & +E[U_{1}^{*}U_{5}]E[U_{2}^{*}U_{6}]E[U_{3}^{*}U_{4}]\nonumber \\
 & +E[U_{1}^{*}U_{6}]E[U_{2}^{*}U_{4}]E[U_{3}^{*}U_{5}]\nonumber \\
 & +E[U_{1}^{*}U_{6}]E[U_{2}^{*}U_{5}]E[U_{3}^{*}U_{4}].\label{eq:iss_6}\end{align}
\\

Let's now start the new proof. We are interested in the PSD $\hat{G}_{x,p}(f)$
of the NLI field $U_{x,p}(L,t)=\mathcal{F}^{-1}[\tilde{U}_{x,p}(L,f)]$.
By theorem 1 we have: \begin{equation}
E[\tilde{U}_{x,p}(L,f)\tilde{U}_{x,p}^{*}(L,u)]=\hat{G}_{x,p}(f)\delta(u-f).\label{eq:uno}\end{equation}

The left hand side can be explicitly calculated using (\ref{eq:field_x}):\begin{align}
 & \frac{E[\tilde{U}_{x,p}(L,f)\tilde{U}_{x,p}^{*}(L,u)]}{\Phi_{NL}^{2}}=\nonumber \\
 & E[\iint_{-\infty}^{\infty}\tilde{\eta}(f_{1}f_{2})[\tilde{U}_{x}(0,f+f_{1})\tilde{U}_{x}^{*}(0,f+f_{1}+f_{2})\tilde{U}_{x}(0,f+f_{2})+\nonumber \\
 & \tilde{U}_{x}(0,f+f_{1})\tilde{U}_{y}^{*}(0,f+f_{1}+f_{2})\tilde{U}_{y}(0,f+f_{2})]\d f_{1}\d f_{2}\cdot\nonumber \\
 & \iint_{-\infty}^{\infty}\tilde{\eta}(f_{3}f_{4})^{*}[\tilde{U}_{x}^{*}(0,u+f_{3})\tilde{U}_{x}(0,u+f_{3}+f_{4})\tilde{U}_{x}^{*}(0,u+f_{4})+\nonumber \\
 & \tilde{U}_{x}^{*}(0,u+f_{3})\tilde{U}_{y}(0,u+f_{3}+f_{4})\tilde{U}_{y}^{*}(0,u+f_{4})]\d f_{3}\d f_{4}]=\nonumber \\
 & \iiiint_{-\infty}^{\infty}\d f_{1}\d f_{2}\d f_{3}\d f_{4}\tilde{\eta}(f_{1}f_{2})\tilde{\eta}(f_{3}f_{4})^{*}\cdot\nonumber \\
 & \{E[\tilde{U}_{x}(0,f+f_{1})\tilde{U}_{x}^{*}(0,f+f_{1}+f_{2})\tilde{U}_{x}(0,f+f_{2})\tilde{U}_{x}^{*}(0,u+f_{3})\tilde{U}_{x}(0,u+f_{3}+f_{4})\tilde{U}_{x}^{*}(0,u+f_{4})]+\nonumber \\
 & E[\tilde{U}_{x}(0,f+f_{1})\tilde{U}_{y}^{*}(0,f+f_{1}+f_{2})\tilde{U}_{y}(0,f+f_{2})\tilde{U}_{x}^{*}(0,u+f_{3})\tilde{U}_{y}(0,u+f_{3}+f_{4})\tilde{U}_{y}^{*}(0,u+f_{4})]+\nonumber \\
 & 2\mbox{Re}(E[\tilde{U}_{x}(0,f+f_{1})\tilde{U}_{x}^{*}(0,f+f_{1}+f_{2})\tilde{U}_{x}(0,f+f_{2})\tilde{U}_{x}^{*}(0,u+f_{3})\tilde{U}_{y}(0,u+f_{3}+f_{4})\tilde{U}_{y}^{*}(0,u+f_{4})])\}.\label{eq:campo_1}\end{align}

Now, putting together Theorems 1 and 2, Appendix 1 shows the following 

\textbf{Theorem 3} 

For jointly stationary circular complex Gaussian zero-mean processes
$A(t),B(t),C(t),D(t),E(t),F(t)$ we have the general formula\begin{align}
E\left[\tilde{A}(f+f_{1})\tilde{B}^{*}(f+f_{1}+f_{2})\tilde{C}(f+f_{2})\tilde{D}^{*}(u+f_{3})\tilde{E}(u+f_{3}+f_{4})\tilde{F}^{*}(u+f_{4})\right] & =\nonumber \\
\Bigl[G_{ab}(f+f_{1})G_{cd}(f)G_{ef}(f+f_{4})\delta(f_{2})\delta(f_{3})+\nonumber \\
G_{ab}(f+f_{1})G_{ed}(f+f_{3})G_{cf}(f)\delta(f_{2})\delta(f_{4})+\nonumber \\
G_{cb}(f+f_{2})G_{ad}(f)G_{ef}(f+f_{4})\delta(f_{1})\delta(f_{3})+\nonumber \\
G_{cb}(f+f_{2})G_{ed}(f+f_{3})G_{af}(f)\delta(f_{1})\delta(f_{4})+\nonumber \\
G_{eb}(f+f_{1}+f_{2})G_{ad}(f+f_{1})G_{cf}(f+f_{2})\delta(f_{3}-f_{1})\delta(f_{4}-f_{2})+\nonumber \\
G_{eb}(f+f_{1}+f_{2})G_{cd}(f+f_{2})G_{af}(f+f_{1})\delta(f_{4}-f_{1})\delta(f_{3}-f_{2})\Bigr]\nonumber \\
\cdot\delta(u-f) & \qquad\square\label{eq:form_compact-1}\end{align}

We next apply the general formula (\ref{eq:form_compact-1}) to the
three expectations in (\ref{eq:campo_1}) to get:

First expectation:\begin{align}
 & E\left[\tilde{U}_{x}(0,f+f_{1})\tilde{U}_{x}^{*}(0,f+f_{1}+f_{2})\tilde{U}_{x}(0,f+f_{2})\tilde{U}_{x}^{*}(0,u+f_{3})\tilde{U}_{x}(0,u+f_{3}+f_{4})\tilde{U}_{x}^{*}(0,u+f_{4})\right]=\nonumber \\
 & \delta(u-f)\Bigl[G_{xx}(f+f_{1})G_{xx}(f)G_{xx}(f+f_{4})\delta(f_{2})\delta(f_{3})+\nonumber \\
 & G_{xx}(f+f_{1})G_{xx}(f+f_{3})G_{xx}(f)\delta(f_{2})\delta(f_{4})+\nonumber \\
 & G_{xx}(f+f_{2})G_{xx}(f)G_{xx}(f+f_{4})\delta(f_{1})\delta(f_{3})+\nonumber \\
 & G_{xx}(f+f_{2})G_{xx}(f+f_{3})G_{xx}(f)\delta(f_{1})\delta(f_{4})+\nonumber \\
 & G_{xx}(f+f_{1}+f_{2})G_{xx}(f+f_{1})G_{xx}(f+f_{2})\delta(f_{3}-f_{1})\delta(f_{4}-f_{2})+\nonumber \\
 & G_{xx}(f+f_{1}+f_{2})G_{xx}(f+f_{2})G_{xx}(f+f_{1})\delta(f_{4}-f_{1})\delta(f_{3}-f_{2})\Bigr]\label{eq:Eform_1}\end{align}
where $G_{xx}\equiv\hat{G}_{x}$. Second expectation:\begin{align}
 & E\left[\tilde{U}_{x}(0,f+f_{1})\tilde{U}_{y}^{*}(0,f+f_{1}+f_{2})\tilde{U}_{y}(0,f+f_{2})\tilde{U}_{x}^{*}(0,u+f_{3})\tilde{U}_{y}(0,u+f_{3}+f_{4})\tilde{U}_{y}^{*}(0,u+f_{4})\right]=\nonumber \\
 & \delta(u-f)\Bigl[G_{xy}(f+f_{1})G_{yx}(f)G_{yy}(f+f_{4})\delta(f_{2})\delta(f_{3})+\nonumber \\
 & G_{xy}(f+f_{1})G_{yx}(f+f_{3})G_{yy}(f)\delta(f_{2})\delta(f_{4})+\nonumber \\
 & G_{yy}(f+f_{2})G_{xx}(f)G_{yy}(f+f_{4})\delta(f_{1})\delta(f_{3})+\nonumber \\
 & G_{yy}(f+f_{2})G_{yx}(f+f_{3})G_{xy}(f)\delta(f_{1})\delta(f_{4})+\nonumber \\
 & G_{yy}(f+f_{1}+f_{2})G_{xx}(f+f_{1})G_{yy}(f+f_{2})\delta(f_{3}-f_{1})\delta(f_{4}-f_{2})+\nonumber \\
 & G_{yy}(f+f_{1}+f_{2})G_{yx}(f+f_{2})G_{xy}(f+f_{1})\delta(f_{4}-f_{1})\delta(f_{3}-f_{2})\Bigr]\label{eq:form_compac}\end{align}
where $G_{yy}\equiv\hat{G}_{y}$, and assuming uncorrelated X and
Y we get\begin{align}
 & E\left[\tilde{U}_{x}(0,f+f_{1})\tilde{U}_{y}^{*}(0,f+f_{1}+f_{2})\tilde{U}_{y}(0,f+f_{2})\tilde{U}_{x}^{*}(0,u+f_{3})\tilde{U}_{y}(0,u+f_{3}+f_{4})\tilde{U}_{y}^{*}(0,u+f_{4})\right]=\nonumber \\
 & \delta(u-f)\Bigl[G_{yy}(f+f_{2})G_{xx}(f)G_{yy}(f+f_{4})\delta(f_{1})\delta(f_{3})+\nonumber \\
 & G_{yy}(f+f_{1}+f_{2})G_{xx}(f+f_{1})G_{yy}(f+f_{2})\delta(f_{3}-f_{1})\delta(f_{4}-f_{2})\Bigr].\label{eq:Eform_2}\end{align}

Third expectation:

\begin{align}
 & E\left[\tilde{U}_{x}(0,f+f_{1})\tilde{U}_{x}^{*}(0,f+f_{1}+f_{2})\tilde{U}_{x}(0,f+f_{2})\tilde{U}_{x}^{*}(0,u+f_{3})\tilde{U}_{y}(0,u+f_{3}+f_{4})\tilde{U}_{y}^{*}(0,u+f_{4})\right]=\nonumber \\
 & \delta(u-f)\Bigl[G_{xx}(f+f_{1})G_{xx}(f)G_{yy}(f+f_{4})\delta(f_{2})\delta(f_{3})+\nonumber \\
 & G_{xx}(f+f_{1})G_{yx}(f+f_{3})G_{xy}(f)\delta(f_{2})\delta(f_{4})+\nonumber \\
 & G_{xx}(f+f_{2})G_{xx}(f)G_{yy}(f+f_{4})\delta(f_{1})\delta(f_{3})+\nonumber \\
 & G_{xx}(f+f_{2})G_{yx}(f+f_{3})G_{xy}(f)\delta(f_{1})\delta(f_{4})+\nonumber \\
 & G_{yx}(f+f_{1}+f_{2})G_{xx}(f+f_{1})G_{xy}(f+f_{2})\delta(f_{3}-f_{1})\delta(f_{4}-f_{2})+\nonumber \\
 & G_{yx}(f+f_{1}+f_{2})G_{xx}(f+f_{2})G_{xy}(f+f_{1})\delta(f_{4}-f_{1})\delta(f_{3}-f_{2})\Bigr]\label{eq:form_compact-3}\end{align}
and assuming uncorrelated X and Y we get\begin{align}
 & E\left[\tilde{U}_{x}(0,f+f_{1})\tilde{U}_{x}^{*}(0,f+f_{1}+f_{2})\tilde{U}_{x}(0,f+f_{2})\tilde{U}_{x}^{*}(0,u+f_{3})\tilde{U}_{y}(0,u+f_{3}+f_{4})\tilde{U}_{y}^{*}(0,u+f_{4})\right]=\nonumber \\
 & \delta(u-f)\Bigl[G_{xx}(f+f_{1})G_{xx}(f)G_{yy}(f+f_{4})\delta(f_{2})\delta(f_{3})+\nonumber \\
 & G_{xx}(f+f_{2})G_{xx}(f)G_{yy}(f+f_{4})\delta(f_{1})\delta(f_{3})\Bigr].\label{eq:Eform_3}\end{align}

Substitution of (\ref{eq:Eform_1}),(\ref{eq:Eform_2}),(\ref{eq:Eform_3})
into (\ref{eq:campo_1}) finally gives \begin{align*}
 & \frac{E[\tilde{U}_{x,p}(L,f)\tilde{U}_{x,p}^{*}(L,u)]}{\Phi_{NL}^{2}}=\delta(u-f)\{\iiiint_{-\infty}^{\infty}\d f_{1}\d f_{2}\d f_{3}\d f_{4}\tilde{\eta}(f_{1}f_{2})\tilde{\eta}(f_{3}f_{4})^{*}\cdot\\
 & \cdot\{\Bigl[\hat{G}_{x}(f+f_{1})\hat{G}_{x}(f)\hat{G}_{x}(f+f_{4})\delta(f_{2})\delta(f_{3})+\hat{G}_{x}(f+f_{1})\hat{G}_{x}(f+f_{3})\hat{G}_{x}(f)\delta(f_{2})\delta(f_{4})+\\
 & \hat{G}_{x}(f+f_{2})\hat{G}_{x}(f)\hat{G}_{x}(f+f_{4})\delta(f_{1})\delta(f_{3})+\hat{G}_{x}(f+f_{2})\hat{G}_{x}(f+f_{3})\hat{G}_{x}(f)\delta(f_{1})\delta(f_{4})+\\
 & \hat{G}_{x}(f+f_{1}+f_{2})\hat{G}_{x}(f+f_{1})\hat{G}_{x}(f+f_{2})\delta(f_{3}-f_{1})\delta(f_{4}-f_{2})+\\
 & \hat{G}_{x}(f+f_{1}+f_{2})\hat{G}_{x}(f+f_{2})\hat{G}_{x}(f+f_{1})\delta(f_{4}-f_{1})\delta(f_{3}-f_{2})\Bigr]+\\
 & \Bigl[\hat{G}_{y}(f+f_{2})\hat{G}_{x}(f)\hat{G}_{y}(f+f_{4})\delta(f_{1})\delta(f_{3})+\hat{G}_{y}(f+f_{1}+f_{2})\hat{G}_{x}(f+f_{1})\hat{G}_{y}(f+f_{2})\delta(f_{3}-f_{1})\delta(f_{4}-f_{2})\Bigr]+\\
 & 2\mbox{Re}(\Bigl[\hat{G}_{x}(f+f_{1})\hat{G}_{x}(f)\hat{G}_{y}(f+f_{4})\delta(f_{2})\delta(f_{3})+\hat{G}_{x}(f+f_{2})\hat{G}_{x}(f)\hat{G}_{y}(f+f_{4})\delta(f_{1})\delta(f_{3})\Bigr])\}\}.\end{align*}

From (\ref{eq:uno}), the term multiplying $\delta(u-f)$ must be
the desired PSD. Each pair of delta removes two integrals, so that
the PSD turns out to be (first two lines above produce first 4 lines,
third line above produces 5th line, 4th line above produces 6th and
7th lines, and final line above produces the last two lines):\begin{align*}
\frac{\hat{G}_{x,p}(f)}{\Phi_{NL}^{2}} & =|\tilde{\eta}(0)|^{2}{\color{green}\iintop_{-\infty}^{\infty}\hat{G}_{x}(f+f_{1})\hat{G}_{x}(f)\hat{G}_{x}(f+f_{4})\d f_{1}\d f_{4}}\\
 & +|\tilde{\eta}(0)|^{2}{\color{green}\iintop_{-\infty}^{\infty}\hat{G}_{x}(f+f_{1})\hat{G}_{x}(f+f_{3})\hat{G}_{x}(f)\d f_{1}\d f_{3}}\\
 & +|\tilde{\eta}(0)|^{2}{\color{green}\iintop_{-\infty}^{\infty}\hat{G}_{x}(f+f_{2})\hat{G}_{x}(f)\hat{G}_{x}(f+f_{4})\d f_{2}\d f_{4}}\\
 & +|\tilde{\eta}(0)|^{2}{\color{green}\iintop_{-\infty}^{\infty}\hat{G}_{x}(f+f_{2})\hat{G}_{x}(f+f_{3})\hat{G}_{x}(f)\d f_{2}\d f_{3}}\\
 & +2\iintop_{-\infty}^{\infty}|\tilde{\eta}(f_{1}f_{2})|^{2}\hat{G}_{x}(f+f_{1}+f_{2})\hat{G}_{x}(f+f_{1})\hat{G}_{x}(f+f_{2})\d f_{1}\d f_{2}\\
 & +|\tilde{\eta}(0)|^{2}{\color{red}\iintop_{-\infty}^{\infty}\hat{G}_{y}(f+f_{2})\hat{G}_{x}(f)\hat{G}_{y}(f+f_{4})\d f_{2}\d f_{4}}\\
 & +\iintop_{-\infty}^{\infty}|\tilde{\eta}(f_{1}f_{2})|^{2}\hat{G}_{y}(f+f_{1}+f_{2})\hat{G}_{x}(f+f_{1})\hat{G}_{y}(f+f_{2})\d f_{1}\d f_{2}\\
 & +2\left({\color{magenta}|\tilde{\eta}(0)|^{2}\iintop_{-\infty}^{\infty}\hat{G}_{x}(f+f_{1})\hat{G}_{x}(f)\hat{G}_{y}(f+f_{4})\d f_{1}\d f_{4}}\right.+\\
 & \,\,\,\,\,\,\left.{\color{magenta}|\tilde{\eta}(0)|^{2}\iintop_{-\infty}^{\infty}\hat{G}_{x}(f+f_{2})\hat{G}_{x}(f)\hat{G}_{y}(f+f_{4})\d f_{2}\d f_{4}}\right).\end{align*}

In summary, considering that by construction $\tilde{\eta}(0)=1$,
we have:\begin{align*}
\frac{\hat{G}_{x,p}(f)}{\Phi_{NL}^{2}} & =2\iintop_{-\infty}^{\infty}|\tilde{\eta}(f_{1}f_{2})|^{2}\hat{G}_{x}(f+f_{1}+f_{2})\hat{G}_{x}(f+f_{1})\hat{G}_{x}(f+f_{2})\d f_{1}\d f_{2}\\
 & +\iintop_{-\infty}^{\infty}|\tilde{\eta}(f_{1}f_{2})|^{2}\hat{G}_{y}(f+f_{1}+f_{2})\hat{G}_{x}(f+f_{1})\hat{G}_{y}(f+f_{2})\d f_{1}\d f_{2}\\
 & +\hat{G}_{x}(f)\left({\color{green}4}(\int_{-\infty}^{\infty}\hat{G}_{x}(f)df)^{2}+{\color{magenta}4}\int_{-\infty}^{\infty}\hat{G}_{x}(f)df\int_{-\infty}^{\infty}\hat{G}_{y}(f)df+{\color{red}1}(\int_{-\infty}^{\infty}\hat{G}_{y}(f)df)^{2}\right)\end{align*}
which confirms Johannisson's equation (\ref{eq:G_xp-2}) and completes
the desired alternative proof.

\section{Conclusions}

We have presented an alternative derivation of Johannisson's result
\cite{Pontus}. We first remark that our new method is able to deal
with correlated X and Y, although this feature was not exploited in
the present paper. Next we note that we did not have to assume independent
input spectral lines: this comes naturally from the stationarity of
the input process. Finally, the truly critical assumption in the model
in \cite{Pontus,carena_JLT} is therefore the assumption of Gaussianity
at any $z$ during propagation, which is implicit in the assumption
of a Gaussian input process, and the fact that the {}``forcing terms''
in the RP equation are the linearly distorted signals at any $z$,
which thus remain Gaussian. 

Therefore the true limit of the model in \cite{Pontus,carena_JLT}
is that indeed starting from a non-Gaussian spectrum such as the one
of a digitally modulated signal%
\footnote{Although the authors in \cite{carena_JLT} present in their Appendix
B an appealing heuristic justification of their Gaussian signal assumption,
still their invoking the central limit theorem at their equation (37)
is\emph{ not rigorous}. They would conclude that any digitally modulated
signal with any number of levels has a Gaussian Fourier transform
(which in turn implies the time-domain signal itself is Gaussian),
which is clearly \emph{not} the case. %
}, it takes some finite propagation in a non-infinite dispersion line
to approximately get both a Gaussian spectrum and a Gaussian-like
time-domain signal.

\section*{Appendix 1}

In this Appendix we prove Theorem 3 in the text. Assuming jointly
stationary circular complex Gaussian zero-mean processes $A(t),B(t),C(t),D(t),E(t),F(t)$,
we have by using (\ref{eq:papoulis}) in (\ref{eq:iss_6}):\begin{align*}
T\triangleq E\left[\tilde{A}(f+f_{1})\tilde{B}^{*}(f+f_{1}+f_{2})\tilde{C}(f+f_{2})\tilde{D}^{*}(u+f_{3})\tilde{E}(u+f_{3}+f_{4})\tilde{F}^{*}(u+f_{4})\right] & =\\
\underbrace{E[\tilde{B}^{*}(f+f_{1}+f_{2})\tilde{A}(f+f_{1})]}_{G_{ab}(f+f_{1})\delta(f_{2})}\underbrace{E[\tilde{D}^{*}(u+f_{3})\tilde{C}(f+f_{2})]}_{G_{cd}(f+f_{2})\delta(u+f_{3}-f-f_{2})}\underbrace{E[\tilde{F}^{*}(u+f_{4})\tilde{E}(u+f_{3}+f_{4})]}_{G_{ef}(u+f_{3}+f_{4})\delta(f_{3})} & +\\
\underbrace{E[\tilde{B}^{*}(f+f_{1}+f_{2})\tilde{A}(f+f_{1})]}_{G_{ab}(f+f_{1})\delta(f_{2})}\underbrace{E[\tilde{D}^{*}(u+f_{3})\tilde{E}(u+f_{3}+f_{4})]}_{G_{ed}(u+f_{3}+f_{4})\delta(f_{4})}\underbrace{E[\tilde{F}^{*}(u+f_{4})\tilde{C}(f+f_{2})]}_{G_{cf}(f+f_{2})\delta(u+f_{4}-f-f_{2})} & +\\
\underbrace{E[\tilde{B}^{*}(f+f_{1}+f_{2})\tilde{C}(f+f_{2})]}_{G_{cb}(f+f_{2})\delta(f_{1})}\underbrace{E[\tilde{D}^{*}(u+f_{3})\tilde{A}(f+f_{1})]}_{G_{ad}(f+f_{1})\delta(u+f_{3}-f-f_{1})}\underbrace{E[\tilde{F}^{*}(u+f_{4})\tilde{E}(u+f_{3}+f_{4})]}_{G_{ef}(u+f_{3}+f_{4})\delta(f_{3})} & +\\
\underbrace{E[\tilde{B}^{*}(f+f_{1}+f_{2})\tilde{C}(f+f_{2})]}_{G_{cb}(f+f_{2})\delta(f_{1})}\underbrace{E[\tilde{D}^{*}(u+f_{3})\tilde{E}(u+f_{3}+f_{4})]}_{G_{ed}(u+f_{3}+f_{4})\delta(f_{4})}\underbrace{E[\tilde{F}^{*}(u+f_{4})\tilde{A}(f+f_{1})]}_{G_{af}(f+f_{1})\delta(u+f_{4}-f-f_{1})} & +\\
\underbrace{E[\tilde{B}^{*}(f+f_{1}+f_{2})\tilde{E}(u+f_{3}+f_{4})]}_{G_{eb}(u+f_{3}+f_{4})\delta(f+f_{1}+f_{2}-u-f_{3}-f_{4})}\underbrace{E[\tilde{D}^{*}(u+f_{3})\tilde{A}(f+f_{1})]}_{G_{ad}(f+f_{1})\delta(u+f_{3}-f-f_{1})}\underbrace{E[\tilde{F}^{*}(u+f_{4})\tilde{C}(f+f_{2})]}_{G_{cf}(f+f_{2})\delta(u+f_{4}-f-f_{2})} & +\\
\underbrace{E[\tilde{B}^{*}(f+f_{1}+f_{2})\tilde{E}(u+f_{3}+f_{4})]}_{G_{eb}(u+f_{3}+f_{4})\delta(f+f_{1}+f_{2}-u-f_{3}-f_{4})}\underbrace{E[\tilde{D}^{*}(u+f_{3})\tilde{C}(f+f_{2})]}_{G_{cd}(f+f_{2})\delta(u+f_{3}-f-f_{2})}\underbrace{E[\tilde{F}^{*}(u+f_{4})\tilde{A}(f+f_{1})]}_{G_{af}(f+f_{1})\delta(u+f_{4}-f-f_{1})} & =\end{align*}
thus\begin{align*}
T & =G_{ab}(f+f_{1})G_{cd}(f+f_{2})G_{ef}(u+f_{3}+f_{4})\delta(f_{2})\delta(f_{3})\delta(u+f_{3}-f-f_{2})\\
 & +G_{ab}(f+f_{1})G_{ed}(u+f_{3}+f_{4})G_{cf}(f+f_{2})\delta(f_{2})\delta(f_{4})\delta(u+f_{4}-f-f_{2})\\
 & +G_{cb}(f+f_{2})G_{ad}(f+f_{1})G_{ef}(u+f_{3}+f_{4})\delta(f_{1})\delta(f_{3})\delta(u+f_{3}-f-f_{1})\\
 & +G_{cb}(f+f_{2})G_{ed}(u+f_{3}+f_{4})G_{af}(f+f_{1})\delta(f_{1})\delta(f_{4})\delta(u+f_{4}-f-f_{1})\\
 & +G_{eb}(u+f_{3}+f_{4})G_{ad}(f+f_{1})G_{cf}(f+f_{2})\cdot\\
 & \cdot\delta(u+f_{4}-f-f_{2})\delta(u+f_{3}-f-f_{1})\delta(f+f_{1}+f_{2}-u-f_{3}-f_{4})\\
 & +G_{eb}(u+f_{3}+f_{4})G_{cd}(f+f_{2})G_{af}(f+f_{1})\cdot\\
 & \cdot\delta(u+f_{4}-f-f_{1})\delta(u+f_{3}-f-f_{2})\delta(f+f_{1}+f_{2}-u-f_{3}-f_{4}).\end{align*}

Now we use the sampling property of the delta to write, e.g. for the
first line where $f_{2}=0$ and $f_{3}=0$, \[
G_{ab}(f+f_{1})G_{cd}(f)G_{ef}(u+f_{4})\delta(f_{2})\delta(f_{3})\delta(u-f)\]
and e.g. for the last line where $u+f_{4}=f+f_{1}$ and $u+f_{3}=f+f_{2}$
which we add up to get\[
{\color{green}u+f_{3}+f_{4}=(f-u)+f+f_{1}+f_{2}}\]
whence\[
{\color{red}f+f_{1}+f_{2}-u-f_{3}-f_{4}}={\color{red}u-f}\]
so that the last line writes as\begin{align*}
 & G_{eb}({\color{green}u+f_{3}+f_{4}})G_{cd}(f+f_{2})G_{af}(f+f_{1})\delta(u+f_{4}-f-f_{1})\delta(u+f_{3}-f-f_{2})\delta({\color{red}f+f_{1}+f_{2}-u-f_{3}-f_{4}})=\\
 & G_{eb}({\color{green}(f-u)+f+f_{1}+f_{2}})G_{cd}(f+f_{2})G_{af}(f+f_{1})\cdot\delta(u+f_{4}-f-f_{1})\delta(u+f_{3}-f-f_{2})\delta({\color{red}u-f})\stackrel{(\mbox{use\,}u=f)}{=}\\
 & G_{eb}({\color{green}f+f_{1}+f_{2}})G_{cd}(f+f_{2})G_{af}(f+f_{1})\delta(f_{4}-f_{1})\delta(f_{3}-f_{2})\delta({\color{red}u-f}).\end{align*}

We therefore get\begin{align*}
T & =G_{ab}(f+f_{1})G_{cd}(f)G_{ef}(f+f_{4})\delta(f_{2})\delta(f_{3})\delta(u-f)\\
 & +G_{ab}(f+f_{1})G_{ed}(f+f_{3})G_{cf}(f)\delta(f_{2})\delta(f_{4})\delta(u-f)\\
 & +G_{cb}(f+f_{2})G_{ad}(f)G_{ef}(f+f_{4})\delta(f_{1})\delta(f_{3})\delta(u-f)\\
 & +G_{cb}(f+f_{2})G_{ed}(f+f_{3})G_{af}(f)\delta(f_{1})\delta(f_{4})\delta(u-f)\\
 & +G_{eb}(f+f_{1}+f_{2})G_{ad}(f+f_{1})G_{cf}(f+f_{2})\delta(f_{4}-f_{2})\delta(f_{3}-f_{1})\delta(u-f)\\
 & +G_{eb}(f+f_{1}+f_{2})G_{cd}(f+f_{2})G_{af}(f+f_{1})\delta(f_{4}-f_{1})\delta(f_{3}-f_{2})\delta(u-f).\end{align*}
whence the final form (\ref{eq:form_compact-1}) given in Theorem
3.
\end{document}